\documentstyle[11pt,IAUS212,twoside,epsf]{article}

\def\edcomment#1{\iffalse\marginpar{\raggedright\sl#1\/}\else\relax\fi}
\reversemarginpar

\begin{document}
\vspace*{1cm}
\title{The feedback of massive stars on the ISM: XMM-Newton view of the LMC
  superbubble N51D}
 \author{Dominik J. Bomans}
\affil{Astronomical Institute of the Ruhr-University Bochum,
  Universit\"atsstr. 150/NA7, D-44780 Bochum, Germany}
\author{J\"orn Rossa}
\affil{Astronomical Institute of the Ruhr-University Bochum,
  Universit\"atsstr. 150/NA7, D-44780 Bochum, Germany}
\author{Kerstin Weis}
\affil{Max-Planck-Institute for Radioastronomy, Auf dem H\"ugel 69, D-53121 
Bonn, Germany}
\author{Konrad Dennerl}
\affil{Max-Planck-Institute for Extraterrestrial Physics, 
Giessenbachstra{\ss}e, D-85748 Garching, Germany}

\begin{abstract}
N51D (= DEM\,L\,192) appears at first glance as a nearly circular, 120\,pc 
diameter bubble of ionized gas around the LMC OB association LH\,54. A deeper
look reveals a complex web of filaments and deviations from radial expansion.
Using a deep XMM-Newton X-ray pointing centered on N51D we find that 
diffuse soft X-ray emitting gas fills the whole superbubble as
delineated by the H$\alpha$ filaments. Contrary to recent findings for 
galactic winds, the correlation between H$\alpha$ and X-ray surface 
brightness is not good. The X-ray spectrum of this diffuse gas cannot be 
fitted with the LMC abundance pattern, but implies an overabundance of at 
least oxygen and neon, consistent with recent enrichment from supernovae 
type II. Some indications for enhanced mixing at the brightest region of the
H$\alpha$ shell and for a beginning outflow of the hot gas were also detected.
\end{abstract}

\section{Introduction}
Feedback of massive stars on the interstellar medium (ISM) comes in
the form of ionizing radiation and mechanical energy input from stellar 
winds and the supernova type II (and Ib and Ic) explosions at the end 
of the stellar live\-span. Importantly, massive stars occur very rarely 
in isolation but form stellar associations and even dense clusters.

To understand the impact of large aggregations of massive stars on
their host galaxies, especially during the phase of galaxy formation,
a sound understanding of the involved processes is needed.
One especially favourable laboratory is the Large Magellanic Cloud
(LMC). Located at high (southern) galactic latitude it offers low 
foreground extinction and \hbox{H\,{\sc i}} absorption, nearly face-on view, 
low intrinsic depth, and therefore small line-of-sight confusion. 
With a distance of 50\,kpc it is nearby enough to use normal
ground-based telescopes for detailed star-by-star analysis of its 
stellar content and a high resolution view of its ISM.

\begin{figure}
\plotfiddle{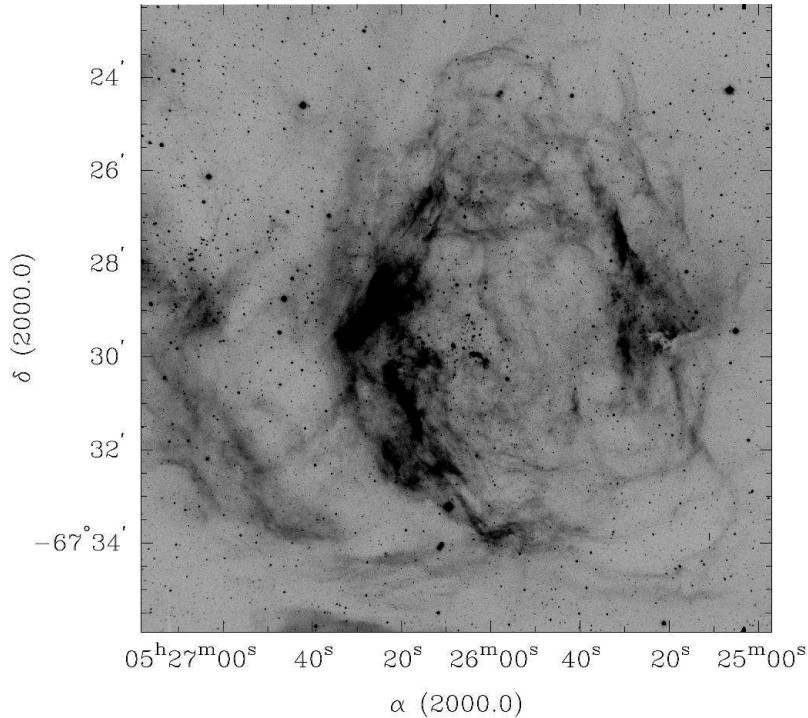}{8.5cm}{0}{54}{54}{-170}{-20}
\caption{H$\alpha$ image of the LMC superbubble N51D at sub-arcsecond 
seeing taken with the CTIO 0.9m telescope.}
\end{figure}

\section{The laboratory N51D}
The superbubble N51D (Henize 1956) (=\,DEM\,L\,192; Davies, Elliot, \&
Meaburn 1976) is a nearly spherical shell with an diameter of about
120\,pc (see Figure~1). The superbubble is located in the southeast corner 
of the supergiant shell LMC4 which encircles Shapley constellation III, a 
superassociation in the north of the LMC. N51D contains two young 
associations (LH\,51 and LH\,54) for which a good amount of spectral 
classification and good photometry is available (e.g.\ Oey \& Smedley 
1998).  

The expansion of the superbubble was studied by Lasker (1980) and
Meaburn \& Terret (1980), who derived an expansion velocity of 35 km
s$^{-1}$ with some indications of different expansion velocity of 
front and back side of the superbubble. N51D is a source of diffuse X-ray 
emission (Chu \& Mac Low 1990), who also found that the eastern side is of 
higher surface brightness. They interpreted this enhanced X-ray surface 
brightness compared to the model predictions for a windblown bubble (Weaver 
et al. 1977) as a recent supernova blast wave hitting the eastern side of 
the shell. More recently, ROSAT data of N51D were presented by Bomans (2001) 
and Dunne, Points, \& Chu (2001).

It is interesting to note here, that absorption of \hbox{N\,{\sc v}} from 
the front side of the superbubble was reported by de Boer \& Nash
(1982) using IUE high dispersion spectra. 

\begin{figure}
\plotfiddle{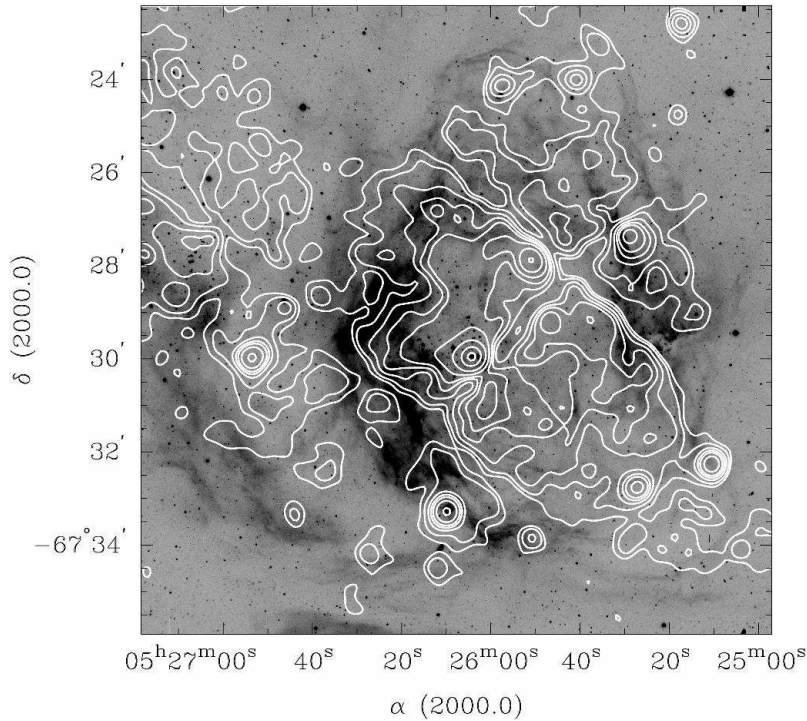}{8.5cm}{0}{54}{54}{-170}{-20}
\caption{Same H$\alpha$ image of the superbubble N51D as in Figure 1, 
overlayed with contours of the 0.3 to 2 keV X-ray emission from a 
subregion of the XMM-Newton EPIC pn image. The vertical contour structures 
(bisectors) are of instrumental origin (borders of the individual CCDs).}
\end{figure}

\section{The hot gas as seen by XMM-Newton}

We chose N51D as a target for a detailed study on the feedback of
massive stars on the interstellar medium using the XMM-Newton 
X-ray observatory. N51D was observed for a total of 32 ksec.   
Data reduction was performed using SAS 5.2 and XSPEC 11.1.
We present here first results using the EPIC pn CCDs. Results from the 
MOS CCDs and the UV and optical imaging using the optical monitor of
XMM-Newton will be presented elsewhere.  

\begin{figure}
\plotfiddle{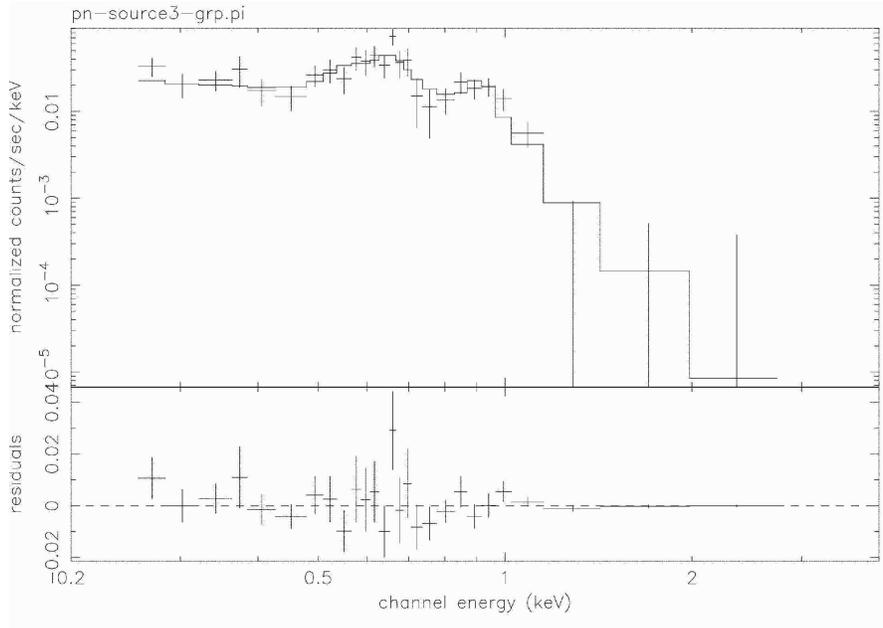}{8.5cm}{0}{60}{60}{-170}{-20}
\caption{XMM-Newton EPIC pn spectrum of a circular region in the high X-ray
surface brightness area inside N51D. The best fitting Mewe-Kaastra-Liedahl
equilibrium plasma model is overplotted.}
\end{figure}

The relation of the soft X-ray photons (0.3 to 2 keV) from the
XMM-Newton pn data with N51D is shown in Figure~2. Besides several 
point sources, diffuse X-ray emission is present all over the optical 
shape of the superbubble. Only the eastern rim of N51D shows enhanced 
surface brightness in X-ray. Apparently we detected the emission of
hot gas filling the superbubble and not solely the interface between 
the hot interior and the cold shell wall, as recently argued for in the case 
of hot outflows of galaxies (e.g.\ Strickland 2002).

We fitted the X-ray spectrum extracted inside the optical boundaries
of N51D with a Mewe-Kaastra-Liedahl equilibrium plasma model (plus a
powerlaw to account for the point sources inside N51D). The foreground 
\hbox{H\,{\sc i}} was first fitted and later fixed to $\log N_{\rm H} = 20.5$,
roughly consistent with results from the ROSAT studies and \hbox{H\,{\sc i}} 
emission profiles in this direction. The plasma temperature is well 
constrained at T $= 0.27 \pm 0.05$ keV, very similar to the ROSAT results 
(Bomans 2001; Dunne, Points, \& Chu 2001). Due to the much higher spectral 
resolution of the XMM-Newton EPIC pn data, coupled with the number of photons 
detected, it is possible to look into considerable detail of the metal 
abundance pattern of the gas. Indeed, an acceptable fit is only possible with 
oxygen and neon abundances enhanced by a factor of 2 over the metal abundance 
of the LMC ISM (e.g.\ Wilcots 1994). The spectrum and the best fitting model 
is plotted in Figure~3. We also selected smaller regions inside N51D (a X-ray 
bright region coinciding with bright H$\alpha$ emission, an X-ray bright 
region with very faint H$\alpha$ emission, and a region with low diffuse X-ray
surface brightness). The spectral fits confirm the results, but also hint a 
lower metal abundance in the region of the bright north-eastern H$\alpha$ 
ridge.

The diffuse X-ray emission north-east of N51D is inside the supergiant 
shell LMC4. The spectrum indicates similar metal enhancement of the
hot LMC gas as reported by Dennerl et al. (2001) for the region 
near SN1987A.

\section{X-ray point sources}
The XMM-Newton field centered on N51D also contains an appreciable
number of point sources, partly intrinsic to the LMC, partly
background AGN and galactic foreground sources. One X-ray source
is especially noteworthy. It coincides exactly with the LMC O8I+WC5
binary Sk-67 104 (= HD36402). The spectrum can be fitted with a
powerlaw, exhibiting a soft spectral index ($\Gamma = 2.6$), while being 
surprisingly luminous (L$_{\rm x} = 9 \times 10^{35}$ erg\,s$^{-1}$). The 
nature of the source is unclear yet and cannot be explained solely by 
colliding winds of the two massive stars.

\section{Conclusions}
Diffuse X-ray emission is present over the whole area of N51D as
defined in H$\alpha$, including the large, faint H$\alpha$ protrusions to 
the north and south-west, hinting at a beginning outflow of the hot gas out 
of the bubble.  

Preliminary spectral analysis of the data from the EPIC pn detectors shows a 
significant overabundance of oxygen and neon over the ISM abundance of the 
LMC. This is consistent with the hot gas being recently enriched by SN type 
II ejecta. Some indications from the spectra also point at enhanced mixing 
in the region of the bright north-eastern H$\alpha$ ridge.  


Clearly the high signal-to-noise X-ray spectra from the N51D  
region show clear signs of metal deposition into the hot phase of 
the ISM and its interaction with the warm and cold gas surrounding the
superbubble. Analysis of the MOS detector data and the UV imaging from the
XMM-Newton optical monitor will further improve the situation and will 
allow a detailed look at the processes at work for the thermal, dynamical, 
and chemical evolution of this region in the LMC.

\acknowledgements

The authors acknowledge financial support by the Deutsches Zentrum f\"ur 
Luft- und Raumfahrt (DLR) through grant 50-OR-0102.

\end{document}